\documentclass[journal]{IEEEtran}

\usepackage{graphicx}
\usepackage{cite}
\usepackage[cmex10]{amsmath}
\usepackage{amssymb}

\DeclareMathAlphabet{\mathbit}{OML}{cmr}{bx}{it}

\DeclareMathOperator{\E}{E}

\DeclareMathOperator{\Erf}{erf}

\DeclareMathOperator{\fieldR}{\mathbb{R}}

\newcommand{\ve}[1]{\boldsymbol{#1}}

\newcommand{\exdi}[2]{\E_{#1} \left[#2\right]}

\renewcommand{\exp}[1]{e^{#1}}

\newcommand{\erf}[1]{\Erf \left(#1\right)}
\newcommand{\erfptwo}[1]{\Erf^2 \left(#1\right)}

\newenvironment{proof}[1][Proof]{\begin{trivlist}
\item[\hskip \labelsep {\bfseries #1}]}{\end{trivlist}}
\newenvironment{corollary}[1][Corollary]{\begin{trivlist}
\item[\hskip \labelsep {\bfseries #1}]}{\end{trivlist}}
\newenvironment{proposition}[1][Proposition]{\begin{trivlist}
\item[\hskip \labelsep {\bfseries #1}]}{\end{trivlist}}

\title{A Lower Bound for the Fisher Information Measure}

\author{Manuel~Stein, Amine~Mezghani, and~Josef~A.~Nossek\\Institute for Circuit Theory and Signal Processing\\Technische Universit\"at M\"unchen, Germany\\
E-mail: manuel.stein@tum.de%
\thanks{\copyright~2014 IEEE. Personal use of this material is permitted. However, permission to use this material for any other purposes must be obtained from the IEEE by sending a request to pubs-permissions@ieee.org.}}

\begin{document}

\maketitle

\begin{abstract}
The problem how to approximately determine the absolute value of the Fisher information measure for a general parametric probabilistic system is considered. Having available the first and second moment of the system output in a parametric form, it is shown that the information measure can be bounded from below through a replacement of the original system by a Gaussian system with equivalent moments. The presented technique is applied to a system of practical importance and the potential quality of the bound is demonstrated. 
\end{abstract}


\IEEEpeerreviewmaketitle

\section{Introduction}
\IEEEPARstart{A}{}ssessing the quality of a given parametric probabilistic system with respect to the inference of it's unknown parameters plays a key role in the analysis and design of signal processing systems. Through the Fisher information measure \cite{Fisher21} \cite{Fisher25}, estimation theory provides a method to characterize the capability of such systems with respect to this problem in a compact way. As the information measure is inversely proportional to the minimum achievable variance with unbiased estimators and exhibits desirable properties like convexity \cite{Stam59} \cite{Cohen68}, it can be used for performance characterization or as a figure of merit when aiming at the optimum system design. However, the application of such rigorous tools to practical problems requires a probabilistic model which accurately captures all underlying effects within the system of interest. Therefore, a crucial point during the conceptual transition from real-world physical systems to theoretical mathematical models is to find a correct analytical representation of their probabilistic nature, i.e. to determine how uncertainty impacts these entities. As superposition characterizes in an appropriate way how noise enters into different practical systems, additive models have become a widely-used assumption among engineers and researchers. If the noise is caused by a sufficiently large number of sources, the central limit theorem can be applied in order to justify an independent additive noise model with Gaussian distribution. Otherwise, using the fact that among all additive noise distributions with fixed variance, the Gaussian one minimizes Fisher information \cite{Boek77} \cite{Stoica11} \cite{Park13}, allows to assess the quality of any probabilistic system with independent additive noise in a conservative way.
\subsection{Motivation}
While mathematical tractability makes models with independent and additive noise attractive, such an assumption must be questioned if one takes into account that  in general probabilistic systems follow non-additive laws. This becomes relevant in practice when considering that a variety of technical systems exhibit severe non-linear characteristics behind their dominant noise sources. Therefore, additive independent noise does not provide an accurate and generic mathematical model for the input-output relation of physical systems. In order to contribute to the scientific understanding of this general class of systems, here we show how to approximately validate the Fisher information measure for an arbitrary parametric probabilistic model in a simple way.
\subsection{Related Works}
An early discussion about the least favorable distributions with respect to location and scale estimation problems is found in \cite{Huber72}. For the generalized form of the Fisher information measure of order $s$, report \cite[pp. 73 ff.]{Boek77} shows that for models with location parameter, the exponential power distribution with fixed $s$-th moment attains minimum Fisher information and cites \cite{Stam59} for a proof of the standard case $s=2$. The work \cite{Zivo97} focuses on minimum Fisher information under fixed higher order moments. The articles \cite{UhrKli95} and \cite{Berch09} analyze the problem under a restriction on the support of the system output. A recent lecture note concerned with minimum Fisher information under a univariate model with independent and additive noise is provided by \cite{Stoica11}, while \cite{Park13} generalizes the result to multivariate problems, i.e. models with independent and additive correlated noise.
\subsection{Contribution}
While the discussion has basically focused on determining the additive noise distribution minimizing the Fisher information under a fixed second moment, to the best of our knowledge, none of the previous works provide a bounding approach for the Fisher information under general models including non-additive systems. The strength of the method presented here lies in it's generality, providing a bound on the Fisher information measure under any probabilistic model and it's simplicity, being based exclusively on the dependency between the first two moments of the system output and the model parameters. This can be useful when analyzing the effects of non-linearities inherent in signal processing systems. The presented bounding technique can be interpreted as a replacement of the original system by an appropriate Gaussian model with equivalent moments and therefore allows to use well-known Gaussian expressions in the estimation theoretic analysis of non-additive parametric probabilistic systems. 
\section{System Model}
For the discussion, we assume access to the random output $y\in\mathcal{Y}$ of a parametric probabilistic system $p_{y}(y;\theta)$ and it's deterministic parameter $\theta\in\Theta$, where $\mathcal{Y}\subseteq\fieldR$ denotes the support of the output $y$ and $\Theta\subseteq\fieldR$ is the parameter set. Further, the first and the second central output moment are available in the form
\begin{align}
\mu(\theta)&=\int_{\mathcal{Y}}  y p_{y}(y;\theta) {\rm{d}}y,\label{prop:first_moment}\\
\sigma^2(\theta)&=\int_{\mathcal{Y}}  \big(y-\mu(\theta)\big)^2 p_{y}(y;\theta) {\rm{d}}y. \label{prop:second_moment}
\end{align}
\section{Fisher Information Bound - Univariate Case}
Given the possibility to observe the system output $y$, following the parameterized probability density function $p_{y}(y;\theta)$, the Fisher information measure \cite{Kay93}
\begin{align}
F(\theta)&=\int_{\mathcal{Y}}  p_{y}(y;\theta) \bigg(\frac{\partial \log{p_{y}(y;\theta)}}{\partial\theta}\bigg)^2 {\rm{d}}y\label{measure:fisher}
\end{align}
is associated with the information carried by realizations of the output $y$ for inference of the unknown parameter $\theta$ since the minimum estimation error which is achievable with an unbiased estimate $\hat{\theta}(y)$ is restricted to
\begin{align}
\exdi{y}{\big(\hat{\theta}(y)-\theta\big)^2}\geq\frac{1}{F(\theta)}.
\end{align}
The goal here is to show that if the derivative $\frac{\partial \mu(\theta)}{\partial\theta}$ exists, the Fisher information measure can be lower bounded by
\begin{align}
F(\theta)&\geq \frac{1}{ \sigma^2(\theta) } \bigg(\frac{\partial \mu(\theta) }{\partial\theta} \bigg)^2.\label{measure:fisher:bound}
\end{align}
\subsection{Proof}
Starting from the definition of the information measure (\ref{measure:fisher}) and using the inequality (\ref{inequality:scalar}) from the appendix,
\begin{align}
F(\theta)&=\int_{\mathcal{Y}}  \bigg(\frac{\partial \log{p_{y}(y;\theta)}}{\partial\theta}  \bigg)^2  p_{y}(y;\theta)  {\rm{d}}y\notag\\
&\geq\frac{\Big( \int_{\mathcal{Y}} \big(y-\mu(\theta)\big) \frac{\partial \log{p_{y}(y;\theta)}}{\partial\theta}p_{y}(y;\theta)dy\Big)^2}{ \int_{\mathcal{Y}}  \big(y-\mu(\theta)\big)^2 p_{y}(y;\theta){\rm{d}}y}\notag\\
&=\frac{\Big( \int_{\mathcal{Y}} \big(y-\mu(\theta)\big) \frac{\partial {p_{y}(y;\theta)}}{\partial\theta}{\rm{d}}y\Big)^2}{ \sigma^2(\theta)} \notag\\
&=\frac{\Big( \int_{\mathcal{Y}} y \frac{\partial {p_{y}(y;\theta)}}{\partial\theta}{\rm{d}}y-\int_{\mathcal{Y}} \mu(\theta) \frac{\partial {p_{y}(y;\theta)}}{\partial\theta}{\rm{d}}y\Big)^2}{ \sigma^2(\theta)} \notag\\
&=\frac{\Big(\frac{\partial }{\partial\theta} \int_{\mathcal{Y}} y p_{y}(y;\theta){\rm{d}}y- \mu(\theta) \frac{\partial }{\partial\theta} \int_{\mathcal{Y}}  p_{y}(y;\theta){\rm{d}}y\Big)^2}{ \sigma^2(\theta)} \label{eq:regularity}\\
&=\frac{1}{ \sigma^2(\theta)} \bigg(\frac{\partial \mu(\theta) }{\partial\theta} \bigg)^2,
\end{align}
allows to bound (\ref{measure:fisher}) in general\footnote{Note that step (\ref{eq:regularity}) necessitates $p_y(y;\theta)$ to fulfill certain regularity conditions \cite{Boek77}. This is the case for most systems of practical relevance.} from below.
\section{Fisher Information Bound - Multivariate Case}
For the general case of a multivariate system output $\ve{y}$ with $\mathcal{Y}\subseteq\fieldR^N$ and multiple parameters $\ve{\theta}$ with $\Theta\subseteq\fieldR^K$, the parametric probabilistic system is written $p_{y}(\ve{y};\ve{\theta})$. The required first and the second output moment are given by
\begin{align}
\ve{\mu}(\ve{\theta})&=\int_{\mathcal{Y}}  \ve{y} p_{y}(\ve{y};\ve{\theta}) {\rm{d}}\ve{y}\label{prop_multi:first_moment}\\
\ve{\Sigma}(\ve{\theta})&=\int_{\mathcal{Y}}  \big(\ve{y}-\ve{\mu}(\ve{\theta})\big)\big(\ve{y}-\ve{\mu}(\ve{\theta})\big)^{\rm T}  p_{y}(\ve{y};\ve{\theta}) {\rm{d}}\ve{y},\label{prop_multi:second_moment}
\end{align}
and the Fisher information measure is defined in matrix form
\begin{align}
\ve{F}(\ve{\theta})&=\int_{\mathcal{Y}}  p_{y}(\ve{y};\ve{\theta}) \bigg(\frac{\partial \log{p_{y}(\ve{y};\ve{\theta})}}{\partial\ve{\theta}}  \bigg)^{\rm T} \frac{\partial \log{p_{y}(\ve{y};\ve{\theta})}}{\partial\ve{\theta}} {\rm{d}}\ve{y}.
\end{align}
The goal now is to show that if the derivative $\frac{\partial \ve{\mu}(\ve{\theta}) }{\partial\ve{\theta}}$ exists,
\begin{align}
\ve{F}(\ve{\theta}) \succeq \bigg(\frac{\partial \ve{\mu}(\ve{\theta}) }{\partial\ve{\theta}} \bigg)^{\rm T} \ve{\Sigma}^{-1}(\ve{\theta}) \bigg(\frac{\partial \ve{\mu}(\ve{\theta}) }{\partial\ve{\theta}} \bigg)\label{measure:fisher:bound:mult}
\end{align}
holds, where with $\ve{A},\ve{B}\in\fieldR^{K\times K}$, the matrix inequality
\begin{align}
\ve{A}\succeq\ve{B},
\end{align}
stands for the fact that
\begin{align}
\ve{x}^{\rm T}(\ve{A}-\ve{B})\ve{x}\geq0\quad\quad\quad\forall \ve{x}\in\fieldR^{K}.
\end{align}
\subsection{Proof}
Inequality (\ref{inequality:vector}) from the appendix is used, such that
\begin{align}
\ve{F}(\ve{\theta}) &= \int_{\mathcal{Y}} \bigg( \frac{\partial \log{p_{{y}}(\ve{y};\ve{\theta})} }{\partial \ve{\theta}} \bigg)^{\rm T} \frac{\partial \log{p_{{y}}(\ve{y};\ve{\theta})} }{\partial \ve{\theta}}   p_{y}(\ve{y};\ve{\theta}) {\rm{d}}\ve{y}\notag\\
&\succeq \int_{\mathcal{Y}} \bigg( \frac{\partial \log{p_{{y}}(\ve{y};\ve{\theta})} }{\partial \ve{\theta}} \bigg)^{\rm T}  \big(\ve{y}-\ve{\mu}(\ve{\theta})\big)^{\rm T} p_{y}(\ve{y};\ve{\theta}) {\rm{d}}\ve{y}\cdot \notag\\
&\phantom{XX}\cdot \ve{\Sigma}^{-1}(\ve{\theta}) \int_{\mathcal{Y}} \big(\ve{y}-\ve{\mu}(\ve{\theta})\big)\frac{\partial \log p_{{y}}(\ve{y};\ve{\theta}) }{\partial \ve{\theta}} p_{y}(\ve{y};\ve{\theta}) {\rm{d}}\ve{y} \notag\\
&= \int_{\mathcal{Y}}  \bigg( \frac{\partial p_{{y}}(\ve{y};\ve{\theta}) }{\partial \ve{\theta}}  \bigg)^{\rm T} \big(\ve{y}-\ve{\mu}(\ve{\theta})\big)^{\rm T} {\rm{d}}\ve{y} \cdot \notag\\
&\phantom{XX}\cdot \ve{\Sigma}^{-1}(\ve{\theta}) \int_{\mathcal{Y}} \big(\ve{y}-\ve{\mu}(\ve{\theta})\big) \frac{\partial p_{{y}}(\ve{y};\ve{\theta}) }{\partial \ve{\theta}} {\rm{d}}\ve{y}\notag\\
&= \bigg( \frac{\partial \ve{\mu}(\ve{\theta})}{\partial \ve{\theta}}  \bigg)^{\rm T}  \ve{\Sigma}^{-1}(\ve{\theta})  \bigg( \frac{\partial \ve{\mu}(\ve{\theta})}{\partial \ve{\theta}} \bigg).
\end{align}
\section{Interpretation - Equivalent Pessimistic System}
For an interpretation of the presented bounding technique, we consider instead of the model $p_{y}(y;\theta)$, the Gaussian system
\begin{align}
q_{y}(y;\theta)=\frac{1}{\sqrt{2\pi \sigma^2(\theta)}} \exp{-\frac{(y-\mu(\theta))^2}{2\sigma^2(\theta)}},\label{dens:GaussZero}
\end{align}
with equivalent first and second moment. If the dependency between $ \sigma^2(\theta)$ and the parameter $\theta$ is ignored and
\begin{align}
\frac{\partial \sigma^2(\theta) }{\partial\theta}=0
\end{align}
is postulated, the Fisher information measure
\begin{align}
\tilde{F}(\theta)&=\int_{\mathcal{Y}}  q_{y}(y;\theta) \bigg(\frac{\partial \log{q_{y}(y;\theta)}}{\partial\theta}  \bigg)^2 {\rm{d}}y
\end{align}
attains the absolute value
\begin{align}
\tilde{F}(\theta)=\frac{1}{ \sigma^2(\theta) } \bigg(\frac{\partial \mu(\theta) }{\partial\theta} \bigg)^2. \label{measure:fisher:gauss:uni}
\end{align}
For the multivariate case, the Gaussian system
\begin{align}
q_{y}(\ve{y};\ve{\theta})=\frac{ \exp{-\frac{1}{2} \big(\ve{y}-\ve{\mu}(\ve{\theta})\big)^{\rm T} \ve{\Sigma}^{-1}(\ve{\theta})\big(\ve{y}-\ve{\mu}(\ve{\theta})\big)} }{(2\pi)^{\frac{N}{2}} \big(\det{\ve{\Sigma}(\ve{\theta})}\big)^{\frac{1}{2}} }\label{dens:GaussZeroMult}
\end{align}
exhibits the Fisher information
\begin{align}
\tilde{\ve{F}}(\ve{\theta})&=\int_{\mathcal{Y}}  q_{y}(\ve{y};\ve{\theta}) \bigg(\frac{\partial \log{q_{y}(\ve{y};\ve{\theta})}}{\partial\ve{\theta}}  \bigg)^{\rm T} \frac{\partial \log{q_{y}(\ve{y};\ve{\theta})}}{\partial\ve{\theta}} {\rm{d}}\ve{y}\notag\\
&= \bigg(\frac{\partial \ve{\mu}(\ve{\theta}) }{\partial\ve{\theta}} \bigg)^{\rm T} \ve{\Sigma}^{-1}(\ve{\theta}) \bigg(\frac{\partial \ve{\mu}(\ve{\theta}) }{\partial\ve{\theta}} \bigg),
\end{align}
when claiming that
\begin{align}
\frac{\partial \ve{\Sigma}(\ve{\theta})}{\partial {\theta}_k}=\ve{0}\label{derivative:covariance:multi}
\end{align} 
for $k=1,\ldots,K$. This shows that the bounding approach can be interpreted as a replacement of the original system $p_{y}(\ve{y};\ve{\theta})$ by an equivalent pessimistic counterpart $q_{y}(\ve{y};\ve{\theta})$, for which the inequality
\begin{align}
\ve{F}(\ve{\theta}) \succeq \tilde{\ve{F}}(\ve{\theta})
\end{align}
holds. So, the Fisher information $\ve{F}(\ve{\theta})$ of the original system $p_{y}(\ve{y};\ve{\theta})$ always dominates the information measure $\tilde{\ve{F}}(\ve{\theta})$ calculated for the equivalent Gaussian system $q_{y}(\ve{y};\ve{\theta})$.
\section{Minimum Fisher Information - A Special Case}
A question which has received attention in the field of statistical signal processing is to specify the worst-case independent additive noise model under a fixed variance \cite{Stoica11} \cite{Park13}. In order to show that our approach includes this relevant special case, consider the multivariate additive system model
\begin{align}
\ve{y}=\ve{s}(\theta)+\ve{\eta},
\end{align}
where the location parameter $\theta\in\fieldR$ modulates the signal $\ve{s}(\theta)\in\fieldR^N$ and $\ve{\eta}\in\fieldR^N$ is a zero-mean independent random process with constant second moment $\ve{R}\in\fieldR^{N \times N}$. As the first two central moments are
\begin{align}
\ve{\mu}(\theta) &= \int_{\mathcal{Y}}  \ve{y} p_{y}(\ve{y};\theta) {\rm{d}}\ve{y}\notag\\
&=\ve{s}(\theta)\\
\ve{\Sigma}&= \int_{\mathcal{Y}}  \big(\ve{y}-\ve{\mu}(\theta)\big) \big(\ve{y}-\ve{\mu}(\theta)\big)^{\rm T}  p_{y}(\ve{y};\theta) {\rm{d}}\ve{y}\notag\\
&=\ve{R},
\end{align}
it follows from (\ref{measure:fisher:bound:mult}), that assuming $\ve{\eta}$ to be normally distributed with constant covariance $\ve{R}$ is the worst-case assumption from an estimation theoretic perspective. However, note that the presented result allows a pessimistic statement about the Fisher information even for the more general case with parametric covariance $\ve{R}(\theta)$. Therefore, it provides a tool for the analysis of a brighter class of systems than those considered in \cite{Stoica11} \cite{Park13}.
\section{Theoretical and Practical Applicability}
For a given system model $p_{y}(\ve{y};\theta)$ with $\mathcal{Y}\subseteq\fieldR^N$, the calculation of the Fisher information measure can be difficult, due to the fact that
\begin{align}
F(\theta)&=\int_{\mathcal{Y}}  p_{y}(\ve{y};\theta) \bigg(\frac{\partial \log{p_{y}(\ve{y};\theta)}}{\partial\theta}\bigg)^2 {\rm{d}}\ve{y}
\end{align}
generally requires the evaluation of an $N$-fold integral. The presented bound $\tilde{F}(\theta)$ has the advantage that the calculation of the two required moments
\begin{align}
\ve{\mu}(\theta) &= \int_{\mathcal{Y}}  \ve{y} p_{y}(\ve{y};\theta) {\rm{d}}\ve{y}\notag\\
\ve{\Sigma}&= \int_{\mathcal{Y}}  \big(\ve{y}-\ve{\mu}(\theta)\big) \big(\ve{y}-\ve{\mu}(\theta)\big)^{\rm T}  p_{y}(\ve{y};\theta) {\rm{d}}\ve{y},
\end{align}
can be performed element-wise, which after marginalization, only requires computing a single and a two-fold integral, respectively. Further, for the case where a mathematical model $p_{y}(\ve{y};\theta)$ is not available, which is a situation frequently encountered in practice, our method allows to estimate the Fisher information by determining the required parametric moments in a calibrated measurement or simulation environment. 
\section{Bounding Quality - Hard-Limiter}
In order to finally show, that the presented procedure has the potential to bound the Fisher information measure in an accurate way, we consider, as an example, a Gaussian signal with mean $\theta\in\fieldR$, which is processed by a hard-limiting device
\begin{align}
y=\operatorname{sign_\alpha}(\theta+\eta),\label{example:system}
\end{align}
where the quantization operation with binary output reads as
\begin{align}
\operatorname{sign_\alpha}(x)=
\begin{cases}
+1& \text{if } x \geq \alpha,\\
-1& \text{if } x < \alpha
\end{cases}
\end{align}
and the additive independent noise term $\eta\in\fieldR$ follows the probability density with unit variance
\begin{align}
p_{\eta}(\eta)=\frac{{\rm e}^{-\frac{\eta^2}{2}}}{\sqrt{2\pi}}.
\end{align}
In this case
\begin{align}
p(y=+1;\theta)&=\int_{-\theta+\alpha}^{\infty}p_{\eta}(\eta) {\rm{d}}\eta\notag\\
&=\frac{1}{2}\bigg(1+\erf{\frac{\theta-\alpha}{\sqrt{2}}}\bigg),\\
p(y=-1;\theta)&=\int_{-\infty}^{-\theta+\alpha}p_{\eta}(\eta) {\rm{d}}\eta\notag\\
&=\frac{1}{2}\bigg(1-\erf{\frac{\theta-\alpha}{\sqrt{2}}}\bigg)
\end{align}
and
\begin{align}
\frac{\partial p(y=+1;\theta)}{\partial \theta}&=\frac{{\rm e}^{- \big(\frac{\theta-\alpha}{\sqrt{2}}\big)^2 }}{\sqrt{2 \pi}},\\
\frac{\partial p(y=-1;\theta)}{\partial \theta}&=-\frac{{\rm e}^{- \big(\frac{\theta-\alpha}{\sqrt{2}}\big)^2 }}{\sqrt{2 \pi}}.
\end{align}
Therefore, the exact Fisher information of the system (\ref{example:system}) is
\begin{align}
F(\theta)&=\int_{\mathcal{Y}}   \frac{1}{p_{y}(y;\theta)}  \bigg(\frac{\partial {p_{y}(y;\theta)}}{\partial\theta}  \bigg)^2  {\rm{d}}y\notag\\
&=  \bigg(\frac{{\rm e}^{- \big(\frac{\theta-\alpha}{\sqrt{2}}\big)^2 }}{\sqrt{ 2 \pi}}   \bigg)^2\Bigg( \frac{2}{1+\erf{\frac{\theta-\alpha}{\sqrt{2}}}} + \frac{2}{1-\erf{\frac{\theta-\alpha}{\sqrt{2}}}} \Bigg)\notag\\ 
&=\frac{2}{\pi}   \frac{{\rm e}^{- (\theta-\alpha)^2 }}{ 1-\erfptwo{\frac{\theta-\alpha}{\sqrt{2}}} } .\label{example:fisher}
\end{align}
In order to apply the information bound (\ref{measure:fisher:bound}), the first moment of the output (\ref{example:system}) is found to be
\begin{align}
\mu(\theta)&=\int_{\mathcal{Y}}  y p_{y}(y;\theta) {\rm{d}}y\notag\\
&=\erf{\frac{\theta-\alpha}{\sqrt{2 }}},
\end{align}
while the second central moment is
\begin{align}
\sigma^2(\theta)&=\int_{\mathcal{Y}}  \bigg(y-\erf{\frac{\theta-\alpha}{\sqrt{2 }}}\bigg)^2 p_{y}(y;\theta) {\rm{d}}y\notag\\
&= 1-\erfptwo{\frac{\theta-\alpha}{\sqrt{2}}}.
\end{align}
With the derivative of the first moment 
\begin{align}
\frac{\partial \mu(\theta)}{\partial \theta}= \sqrt{\frac{2}{\pi}} {\rm e}^{- \big(\frac{\theta-\alpha}{\sqrt{2}}\big)^2 },
\end{align}
the lower bound of the Fisher information measure
\begin{align}
\tilde{F}(\theta)&=\frac{1}{ \sigma^2(\theta) } \bigg(\frac{\partial \mu(\theta) }{\partial\theta} \bigg)^2\notag\\
&=\frac{2}{\pi}   \frac{{\rm e}^{- (\theta-\alpha)^2 }}{ 1-\erfptwo{\frac{\theta-\alpha}{\sqrt{2}}} }
\end{align}
interestingly matches the exact result (\ref{example:fisher}).
\section{Conclusion}
We have established a lower bound for the Fisher information measure under an arbitrary parametric probabilistic system. Being able to characterize the first and the second central moment of the system output as functions of the system parameters, consultation of an equivalent Gaussian system leads to a pessimistic description of the information measure. Using the presented approach for a non-linear system of practical relevance, we have demonstrated that the presented bound has the potential to be tight to the exact information measure. In particular, in situations where the analytical characterization of the information measure with the model of interest is difficult, but where the first two central moments are available, the presented technique allows to analyze the system with respect to the achievable estimation performance.
\appendices
\section{}
\begin{proposition}Given multivariate random variables $\ve{x}, \ve{y}\in\fieldR^N$ which follow the joint probability distribution $p_{xy}(\ve{x}, \ve{y})$,
\begin{align}
\exdi{y}{\ve{y}\ve{y}^{\rm T}}\succeq\exdi{xy}{\ve{y}\ve{x}^{\rm T}} \exdi{x}{\ve{x}\ve{x}^{\rm T}}^{-1} \exdi{xy}{\ve{x}\ve{y}^{\rm T}}.\label{inequality:vector}
\end{align}
\begin{proof}
\noindent Given $\ve{x}$, construct the auxiliary random variable
\begin{align}
\ve{\hat{y}}(\ve{x})=\exdi{xy}{\ve{y}\ve{x}^{\rm T}} \exdi{x}{\ve{x}\ve{x}^{\rm T}}^{-1} \ve{x}.
\end{align}
Observing that by construction,
\begin{align}
\exdi{xy}{(\ve{y}-\ve{\hat{y}}(\ve{x})) (\ve{y}-\ve{\hat{y}}(\ve{x}))^{\rm T}} \succeq \ve{0}
\end{align}
proves that
\begin{align}
\exdi{y}{\ve{y}\ve{y}^{\rm T}}-\exdi{xy}{\ve{y}\ve{x}^{\rm T}} \exdi{x}{\ve{x}\ve{x}^{\rm T}}^{-1} \exdi{xy}{\ve{x}\ve{y}^{\rm T}}\succeq\ve{0}.
\end{align}
\end{proof}
\end{proposition}
\begin{corollary} 
Given scalar random variables $x, y\in\fieldR$, which follow the joint probability distribution $p_{xy}(x,y)$, it holds that
\begin{align}
\int_{\mathcal{Y}}y^2 p_{y}(y){\rm{d}}y \int_{\mathcal{X}}x^2 p_{x}(x){\rm{d}}x\geq \bigg(\int_{\mathcal{X}} \int_{\mathcal{Y}} xy p_{xy}(x,y) {\rm{d}}x {\rm{d}}y\bigg)^2.\label{inequality:scalar}
\end{align}
\begin{proof}
With $N=1$, inequality (\ref{inequality:vector}) simplifies to
\begin{align}
\int_{\mathcal{Y}} y^2 p_{y}(y) {\rm{d}}y \geq \frac{\big(\int_{\mathcal{X}} \int_{\mathcal{Y}}   xy p_{xy}(x,y) {\rm{d}}x {\rm{d}}y\big)^2}{ \int_{\mathcal{X}} x^2 p_{x}(x) {\rm{d}}x},
\end{align}
such that (\ref{inequality:scalar}) follows directly.
\end{proof}
\end{corollary}
\bibliographystyle{IEEEbib}

\end{document}